\documentclass[a4paper]{jpconf}
\usepackage{graphicx}

\usepackage{cite}

\begin{document}
\title{Study of the $\eta$ meson production with the polarised proton beam}

\author{P Moskal, M Hodana}

\address{Institute of Physics, Jagiellonian University, PL-30059 Cracow, Poland}

 \ead{p.moskal@uj.edu.pl}

\begin{abstract}
The azimuthally symmetric WASA detector and the polarised proton beam of COSY,
enable investigations of energy and angular dependence of the beam analysing power in the $pp \rightarrow pp\eta$
reaction. The aim of the studies is the determination of the partial wave contributions via interference
terms which are inaccessible from spin averaged observables. The partial wave decomposition is mandatory for the
understanding of the reaction dynamics and for the determination of the $\eta$-proton interaction, independently of
the theoretical paradigm used. 
\end{abstract}

\section{Introduction}
In the low energy regime of Quantum Chromodynamics,
the interaction between quarks and gluons
cannot be treated perturbatively and so far the  understanding  of the processes
governed by the strong forces is unsatisfactory.
Therefore,
it is essential to carry out measurements involving the production and decay
of hadrons and to interpret 
them in the framework of effective field theories
experiencing recently 
an enormous development 
in applications to the description of meson decays and production.
In this contribution we concentrate on the $\eta$ meson.
The progress in understanding of the 
production processes of the $\eta$
meson will strongly rely on the precise 
determination of spin and isospin observables.

\section{Partial Waves}

Independently of the theoretical framework used,
for an unambiguous understanding of the production process  relative magnitudes
from the partial waves contributions must be well established.
This may be achieved by the measurement of the analysing 
power which would enable to perform the
partial wave decomposition  with an accuracy by far better than resulting from the
measurements of the distributions of the spin averaged
cross sections.
This is because the polarisation observables can probe the interference
terms between various partial amplitudes, even if they
are negligible
for the spin averaged distributions.
More importantly,
in the case 
of the $pp\to ppX$ reaction, as
pointed out in reference~\cite{deloff,colinpriv},
the interference terms between the transitions with odd and even
values of the angular momentum of the final state baryons are
bound to vanish for the cross sections.
This characteristic is due to the invariance of all observables under
the exchange of identical nucleons in the final state.
Due to the same reason there is no interference between s and p-waves
of the $\eta$ meson in the differential cross sections~\cite{colinpriv}.
However, s-p interference does not vanish for the proton analysing power,
and thus the precise measurements of $A_y$ could provide
the first determination of the comparatively 
small p-wave contribution~\cite{colinpriv},
unreachable from spin averaged observables.

In the last decade a vast set of the unpolarised observables
has been established at the facilities CELSIUS, COSY and SATURNE 
for $\eta$ production in the collision of nucleons.
The data comprise in principle a lot of interesting information 
concerning the production mechanism and the $\eta$-nucleon interaction.
These, however, could have not been derived unambiguously due to lack
of the knowledge about the relative contributions from the partial 
waves involved.

\section{Presently avaiable data}
The present poor data base of the polarisation observables only allows for 
qualitative conclusions.  

 In order to establish  quantitatively contributions 
from various production processes and to determine 
possible interference terms  more precise measurments of the 
spin observables and more support from the theoretical side is 
needed.
\begin{figure}[h]
\begin{center}
\includegraphics[width=10.5cm]{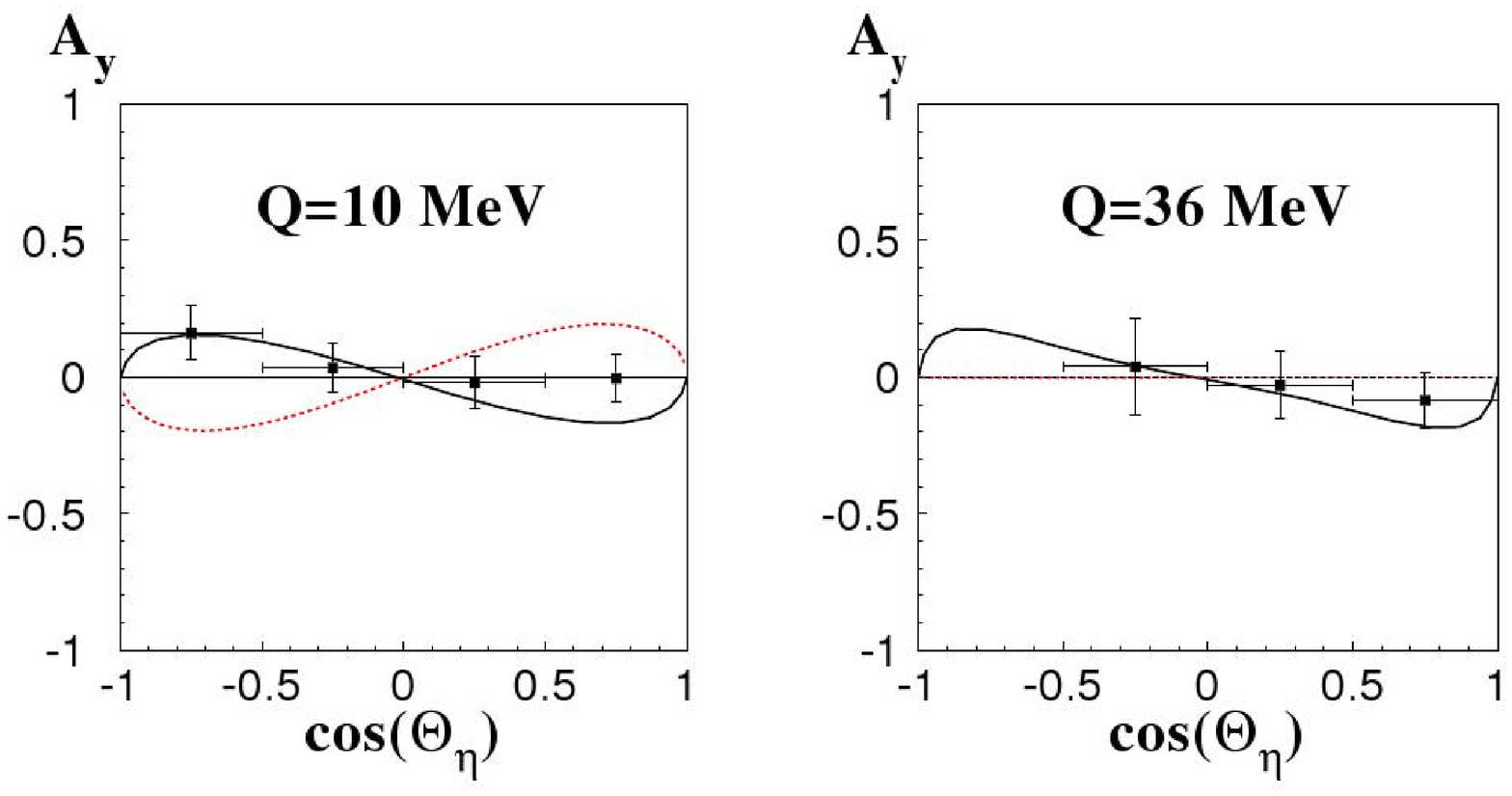}
\end{center}
\vspace{-0.4cm}
\caption{Analysing powers for the $\vec{p}p\to pp\eta$ reaction as functions of $\cos{\theta_{\eta}}$ for
        Q~=~10~MeV (left panel) and Q~=~36~MeV (right panel) obtained by the COSY-11 collaboration~\cite{aycosy11,czyzyk-phd}. 
        Full lines are the
        predictions based on the pseudoscalar meson exchange model~\cite{nakayama}
        whereas the dotted lines represent the 
        calculations based on the vector meson exchange model~\cite{wilkin}.
	       Shown are the statistical uncertainties solely. 
\label{aycosy11}}
\vspace{-0.2cm}
\end{figure}
\vspace{-0.2cm}
\begin{figure}[h]
\begin{center}
\includegraphics[width=7.2cm, height=9cm]{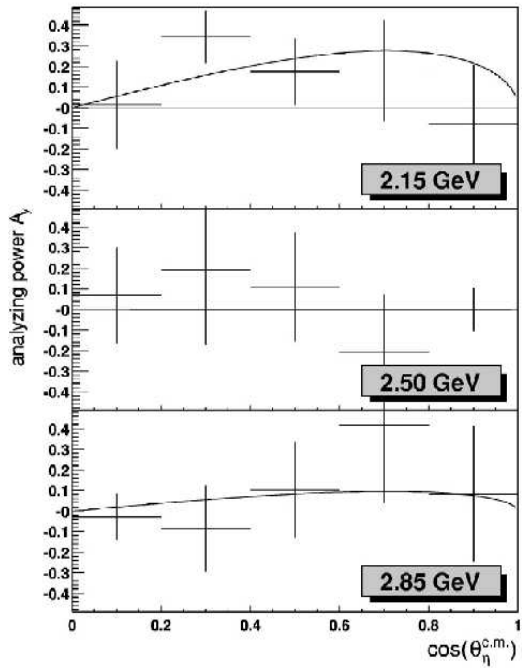}
\end{center}
\vspace{-0.4cm}
\caption{Analysing power  for the $\vec{p}p\to pp\eta$ 
reaction as functions of $\cos{\theta_{\eta}}$ for
        Q~=~324~MeV (upper panel),  Q~=~412~MeV (middle panel)
        and Q~=~554~MeV (lower panel) obtained by the DISTO collaboration~\cite{disto}. 
		The solid curves are fits to the data
        using a formula  based on $\rho$ meson exchange~\cite{wilkin}.
        The figure is adapted from~\cite{disto}.
	\label{aydisto}}
\vspace{-0.2cm}
\end{figure}

So far measurements of the analysing power for the $\vec{p}p\to pp\eta$
reaction have been performed 
in the near threshold energy region at 
excess energies of Q~=~10 and 36~MeV at COSY 
 by the COSY-11 collaboration~\cite{aycosy11,pwinter}, 
and at higher energies for Q~=~324~MeV,  Q~=~412~MeV
and Q~=~554~MeV at SATURNE by the DISTO collaboration~\cite{disto}.
For all studied energies, the determined analysing power is essentially consistent with zero 
implying that the $\eta$ meson is produced predominantly in $s$-wave. 
The achieved statistics, as shown in figures~\ref{aycosy11} and~\ref{aydisto},
 allowed for only a rough determination of the angular dependence
with four and five bins and the errors of A$_y$ equal to $\pm$0.1 and $\pm$0.2, respectively.

Using the WASA-at-COSY facility we intend to determine the energy 
and angular dependence of A$_y$(Q,$\theta$) and the total and differential 
cross sections for the $pp\to pp\eta$ reaction in the excess energy range 
from the threshold up to 100~MeV. In November 2010 first measurement for Q~=~15~MeV 
and Q~=~72~MeV have been conducted~\cite{prop10} .

\section{Dynamics of the $ \mathbf{pp\to pp\eta}$ reaction}
Precise data sets~\cite{chiavassa,calen1,calen2,hibou,smyrski,bergdolt,abdel,hab}
  on the total
  cross section of $\eta$ meson production
  in the $pp\to pp\eta$ reaction allowed to conclude that the reaction proceeds
  through the excitation of one of the protons to the $S_{11}(1535)$ state which
  subsequently deexcites via emission of the $\eta$ meson.
  The crucial observations
  were a large value of the absolute cross section (forty times larger than for the
  $\eta^\prime$ meson) and  isotropic distributions~\cite{hab,Abd01,pm04} of the
  $\eta$ meson emission angle in the reaction center-of-mass system.
More constraints to the theoretical
models~\cite{nakayama,wilkin,germond,laget,moalem,vetter,oset,bati}
have been deduced from the determination of the dependence on the 
isospin of the colliding nucleons~\cite{wasa1}.
The experiments performed by the WASA/PROMICE collaboration~\cite{wasa1}
revealed a strong isospin dependence confirmed at threshold by the COSY-11 group~\cite{pm06,pm09}.
All~together, the confrontation of predictions based upon different
scenarios, involving exchanges of various mesons,  with
the so far determined unpolarised observables and with
the first results on the analysing power
indicate the dominance of $\pi$ exchange 
in the production process~\cite{aycosy11}.
This conclusion is in line  with  the predictions of Nakayama et al.~\cite{nakayama}
and also recent calculations of Shyam~\cite{shyam}.
Yet, the implications seem to be contra-intuitive due to the very large momentum
transfer between the interacting nucleons needed to create the $\eta$ meson
near threshold.
A poor data base of polarisation observables allows, however, only for
qualitative conclusions.
 To establish  quantitatively contributions
from various production processes and to determine
possible interference terms  more precise measurements of the
spin observables
are needed.
Another very interesting feature of $pp\to pp\eta$ reaction is the difficulty 
in reproducing the $pp$ invariant mass distributions~\cite{abdel, pm04, christian, pm10}.
Calculations which include $NN$ FSI and $N\eta$ FSI 
do not match existing data \cite{pm04}. 
To explain the unexpected shape of the distribution, 
 possibility of higher partial-waves is considered. 
Taking into account a $P$-wave contribution one could 
reproduce the $pp$ invariant mass distribution but not the close to threshold cross section dependencies \cite{Nakayama:2003pw}.
To solve this discrepancy, a 
$D_{13}$ resonance has been included~\cite{kanzoMENU}. 
In the calculations, however, the data collected so far are insufficient for the unambiguous extraction of the $S$-wave or $P$-wave contributions.

High statistics data collected by the WASA-at-COSY detector should enable a study of the evolution 
of the analysing power as a function of the invariant mass spectra of the two particle subsystems.
This would shed a light on the still not explained  origin of  structures
in the invariant mass distributions observed independently by the TOF~\cite{abdel}, 
COSY-11~\cite{pm04,pm10}, and CELSIUS/WASA~\cite{christian} collaborations.
It is worth to stress that similar shapes of the invariant mass distributions have
been also observed recently in the case of the $\eta^{\prime}$ meson~\cite{c11klaja}.
In both the $\eta$ and the $\eta^{\prime}$ case the intricate structure remains so far unexplained.

\section{ Studies of $A_y$ with the WASA-at-COSY detector}

For the measurements of the beam analysing power of the $\vec{p}p \to pp\eta$ reaction
we use the axially symmetric WASA-at-COSY
experimental setup~\cite{WaC04}
working as an internal target facility at the cooler synchrotron
COSY~\cite{meier,prasuhn}.
A vertically polarised proton
beam~\cite{stockhorst},
is stored and
accelerated in the COSY ring.
The direction of the polarisation is flipped
from cycle to cycle.
The beam of hydrogen pellets cross the circulating COSY beam
in the center of the WASA detector.
Protons from the $pp\to pp\eta$  reaction
are registered in the Forward Detector
and the gamma quanta  from the $\eta$ meson decay
are detected in the electromagnetic calorimeter.
Both the  invariant mass of the decay products
and the missing mass to the outgoing protons are used for the identification of the $\eta$ meson.
The determination of the beam polarisation and the
control of the systematics  is achieved
by measuring the asymmetries
for  elastically scattered protons for which precise
values of the analysing powers are available~\cite{altmeier}.
The accuracy of these results is 1.2\% and will allow to
control the systematic error of the polarisation
determination to about 1\%.


Based on the online monitoring of the first measurement of the $pp \rightarrow pp\eta$ reaction performed with WASA-at-COSY, the beam polarization achieved in this experiment amount to $\sim70\%$ and $\sim60\%$, for measurements with ~Q=~15~MeV and Q~=~72~MeV, respectively. 

\ack
The work was partially supported by the European Commission
through the \emph{Research Infrastructures} action of the 
\emph{Capacities} Programme:       
Call: FP7-INFRASTRUCTURES-2008-1, Grant Agreement N. 227431, 
by the PrimeNet, by the Polish Ministry of Science and Higher 
Education through grant No. 1202/DFG/2007/03,
by the German Research Foundation (DFG), 
by the FFE grants from the Research Center J{\"u}lich, 
and by the virtual institute \emph{Spin and strong QCD} (VH-VP-231)

\section*{References}
\bibliography{AyBib}

\providecommand{\newblock}{}
\begin{thebibliography}{10}
\expandafter\ifx\csname url\endcsname\relax
  \def\url#1{{\tt #1}}\fi
\expandafter\ifx\csname urlprefix\endcsname\relax\def\urlprefix{URL }\fi
\providecommand{\eprint}[2][]{\url{#2}}

\bibitem{deloff}
Deloff A 2004 {\em Phys. Rev.\/} {\bf C69} 035206 (\textit{Preprint}
  \eprint{nucl-th/0309059})

\bibitem{colinpriv}
Wilkin C 2007 private communication

\bibitem{aycosy11}
Czyzykiewicz R {\em et~al.\/} 2007 {\em Phys. Rev. Lett.\/} {\bf 98} 122003
  (\textit{Preprint} \eprint{hep-ex/0611015})

\bibitem{czyzyk-phd}
Czyzykiewicz R 2006 {\em {Study of the production mechanism of the eta meson in
  proton-proton collisions by means of analysing power measurements}\/} Ph.D.
  thesis (\textit{Preprint} \eprint{nucl-ex/0702010})

\bibitem{nakayama}
Nakayama K, Speth J and Lee T~S~H 2002 {\em Phys. Rev.\/} {\bf C65} 045210
  (\textit{Preprint} \eprint{nucl-th/0202012})

\bibitem{wilkin}
Faldt G and Wilkin C 2001 {\em Phys. Scripta\/} {\bf 64} 427--438
  (\textit{Preprint} \eprint{nucl-th/0104081})

\bibitem{disto}
Balestra F {\em et~al.\/} 2004 {\em Phys. Rev.\/} {\bf C69} 064003

\bibitem{pwinter}
Winter P {\em et~al.\/} 2003 {\em Eur. Phys. J.\/} {\bf A18} 355--357
  (\textit{Preprint} \eprint{nucl-ex/0302010})

\bibitem{prop10}
{Moskal, P and Hodana, M and Cal\'{e}n} H 2010  {Institut f\"ur Kernphysik/COSY
  proposal 185.1}

\bibitem{chiavassa}
Chiavassa E {\em et~al.\/} 1994 {\em Phys. Lett.\/} {\bf B322} 270--274

\bibitem{calen1}
Calen H {\em et~al.\/} 1996 {\em Phys. Lett.\/} {\bf B366} 39--43

\bibitem{calen2}
Calen H {\em et~al.\/} 1997 {\em Phys. Rev. Lett.\/} {\bf 79} 2642--2645

\bibitem{hibou}
Hibou F {\em et~al.\/} 1998 {\em Phys. Lett.\/} {\bf B438} 41--46
  (\textit{Preprint} \eprint{nucl-ex/9802002})

\bibitem{smyrski}
Smyrski J {\em et~al.\/} 2000 {\em Phys. Lett.\/} {\bf B474} 182--187
  (\textit{Preprint} \eprint{nucl-ex/9912011})

\bibitem{bergdolt}
Bergdolt A~M {\em et~al.\/} 1993 {\em Phys. Rev.\/} {\bf D48} 2969--2973

\bibitem{abdel}
Abdel-Bary M {\em et~al.\/} (COSY-TOF) 2003 {\em Eur. Phys. J.\/} {\bf A16}
  127--137 (\textit{Preprint} \eprint{nucl-ex/0205016})

\bibitem{hab}
Moskal P 2004  {Habilitation Dissertation No. 364, Jagellonian University
  Press} (\textit{Preprint} \eprint{hep-ph/0408162})

\bibitem{Abd01}
Abd El-Samad S {\em et~al.\/} (COSY-TOF) 2001 {\em Phys. Lett.\/} {\bf B522}
  16--21 (\textit{Preprint} \eprint{nucl-ex/0107009})

\bibitem{pm04}
Moskal P {\em et~al.\/} 2004 {\em Phys. Rev.\/} {\bf C69} 025203
  (\textit{Preprint} \eprint{nucl-ex/0307005})

\bibitem{germond}
Germond J~F and Wilkin C 1990 {\em Nucl. Phys.\/} {\bf A518} 308--316

\bibitem{laget}
Laget J~M, Wellers F and Lecolley J~F 1991 {\em Phys. Lett.\/} {\bf B257}
  254--258

\bibitem{moalem}
Moalem A, Gedalin E, Razdolskaya L and Shorer Z 1996 {\em Nucl. Phys.\/} {\bf
  A600} 445--460

\bibitem{vetter}
Vetter T, Engel A, Biro T and Mosel U 1991 {\em Phys. Lett.\/} {\bf B263}
  153--156

\bibitem{oset}
Lopez~Alvaredo B and Oset E 1994 {\em Phys. Lett.\/} {\bf B324} 125--129

\bibitem{bati}
Batinic M, Svarc A and Lee T~S~H 1997 {\em Phys. Scripta\/} {\bf 56} 321--324
  (\textit{Preprint} \eprint{nucl-th/9604043})

\bibitem{wasa1}
Calen H {\em et~al.\/} 1998 {\em Phys. Rev.\/} {\bf C58} 2667--2670

\bibitem{pm06}
Moskal P {\em et~al.\/} 2006 {\em J. Phys.\/} {\bf G32} 629--641
  (\textit{Preprint} \eprint{nucl-ex/0507033})

\bibitem{pm09}
Moskal P {\em et~al.\/} 2009 {\em Phys. Rev.\/} {\bf C79} 015208
  (\textit{Preprint} \eprint{0807.0722})

\bibitem{shyam}
Shyam R 2007 {\em Phys. Rev.\/} {\bf C75} 055201 (\textit{Preprint}
  \eprint{nucl-th/0701011})

\bibitem{christian}
Pauly C  In the Proceedings of 11th International Conference on Meson-Nucleon
  Physics and the Structure of the Nucleon (MENU 2007), Julich, Germany, 10-14
  Sep 2007

\bibitem{pm10}
Moskal P {\em et~al.\/} 2010 {\em Eur. Phys. J.\/} {\bf A43} 131--136
  (\textit{Preprint} \eprint{0912.1592})

\bibitem{Nakayama:2003pw}
Nakayama K, Haidenbauer J, Hanhart C and Speth J 2003 {\em Phys. Rev.\/} {\bf
  C68} 045201 (\textit{Preprint} \eprint{nucl-th/0302061})

\bibitem{kanzoMENU}
Nakayama K, Oh Y and Haberzettl H  In the Proceedings of 11th International
  Conference on Meson-Nucleon Physics and the Structure of the Nucleon (MENU
  2007), Julich, Germany, 10-14 Sep 2007

\bibitem{c11klaja}
Klaja P {\em et~al.\/} 2010 {\em Phys. Lett.\/} {\bf B684} 11--16
  (\textit{Preprint} \eprint{1001.5174})

\bibitem{WaC04}
Adam H~H {\em et~al.\/} (WASA-at-COSY) 2004  (\textit{Preprint}
  \eprint{nucl-ex/0411038})

\bibitem{meier}
Maier R 1997 {\em Nucl. Instrum. Meth.\/} {\bf A390} 1--8

\bibitem{prasuhn}
Prasuhn D {\em et~al.\/} 2000 {\em Nucl. Instrum. Meth.\/} {\bf A441} 167--174

\bibitem{stockhorst}
Stockhorst H 2004  (\textit{Preprint} \eprint{physics/0411148})

\bibitem{altmeier}
Altmeier M {\em et~al.\/} (EDDA) 2000 {\em Phys. Rev. Lett.\/} {\bf 85}
  1819--1822

\end{thebibliography}

\end{document}